# Highly tunable ultra-narrow-resonances with optical nano-antenna phased arrays in the infrared

Shi-Qiang Li*[1], Wei Zhou[1], Peijun Guo[1], D. Bruce Buchholz[1], Ziwei Qiu[1], John B. Ketterson[2,3], Leonidas E. Ocola[4], Kazuaki Sakoda[3,5], Robert P. H. Chang[1,3]

[1]Department of Materials Science and Engineering, Northwestern University, 2220 Campus Dr., Evanston, IL 60208-3108, USA;
[2]Department of Physics, Northwestern University, 2145 Sheridan Rd., Evanston, IL 60208-3113, USA;
[3]NU-NIMS Materials Innovation Center, 2220 Campus Dr., Evanston, IL 60208-3108, USA;
[4]Center for Nanoscale Materials, Argonne National Laboratory, 9700 S Cass Ave, Lemont, IL 60439, USA;
[5]National Institute for Materials Science, 1-2-1 Sengen, Tsukuba, Ibaraki 305-0047, Japan

## ABSTRACT

We report our recent development in pursuing high Quality-Factor (high-Q factor) plasmonic resonances, with vertically aligned two dimensional (2-D) periodic nanorod arrays. The 2-D vertically aligned nano-antenna array can have high-Q resonances varying arbitrarily from near infrared to terahertz regime, as the antenna resonances of the nanorod are highly tunable through material properties, the length of the nanorod, and the orthogonal polarization direction with respect to the lattice surface,. The high-Q in combination with the small optical mode volume gives a very high Purcell factor, which could potentially be applied to various enhanced nonlinear photonics or optoelectronic devices. The 'hot spots' around the nanorods can be easily harvested as no index-matching is necessary. The resonances maintain their high-Q factor with the change of the environmental refractive index, which is of great interest for molecular sensing.

**Keywords:** Monopole optical antenna, diffractive coupling, Purcell factor, optical focusing, discrete dipole coupling.

## 1. INTRODUCTION

With the ever-increasing interest to apply plasmonic resonances to various applications, many unusual phenomena have been observed and explained. In the examples such as surface enhanced infrared/Raman spectroscopy[1, 2] and Surface-Plasmon-Amplified-Stimulated-Emission-by-Radiation (SPASER),[3] a common goal is to engineer the resonances so that they can be tuned to different wavelengths without significant change of their quality factors (Q) and field enhancement factors. However, for many plasmonic resonators, high Q factor can only be achieved with index matching technique, which greatly limits the accessibility of the "hot spots" generated around the plasmonic structures. In our recent report, we have demonstrated that a monopole phased array possesses index-matching-free high-Q resonances.[4] Here, we discuss its analytical model in details and how we achieve index-matching-free sharp resonances with large tunability in the phased array system.

*Address now: Electrical and Electronic Engineering, University of Melbourne, Victoria, 3010, Australia; shiqiang.li@unimelb.edu.au; phone +61 0383443819

## 2. RESULTS AND DISCUSSIONS

The analytical model used in this work is an extension of Markel-Schatz Model developed to study the dipole coupling in a periodic array of dipoles.[5-7] The assumptions made in the model are as follows. 1. The entities in the array are small enough comparing to the wavelength of light so that they can be approximated as point dipoles. 2. The array is periodic and the size of the array is large enough so that the boundary condition is negligible. 3. The dipoles are separated far apart so that only dipole field is to be considered.

In the dipole coupling model, there is a very interesting regime - when the dipoles are separated apart at the length scale of the wavelength of light. At this regime, phase interference from the dipole radiation can generate photonic modes. These photonic modes can interact with the dipole resonances located at around the same spectral region; the result of the interaction is an asymmetric spectral lineshape, which could have the resonance strength much greater than either of the interfering resonances. The photonic modes have very sharp linewidth but low strength, while the dipole resonances (such as plasmonic resonances) are broad and strong. The interference between these two types of modes is described in Fano's study on atomic resonances, which is then known as Fano resonances.[8] This interaction between the sharp photonic mode and the broad dipole mode is depicted in Figure 1.

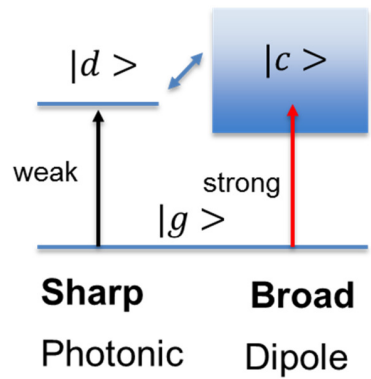

Figure 1. A schematic diagram showing the interaction between the photonic resonance and plasmonic resonance with energy-level representation. As photonic mode in dipole coupling is a sharp mode with very small linewidth, it can be represented as excitation of a particle from ground level |$g$> to an excited level |$d$>. On the other hand, the dipole resonance (such as plasmonic resonance) can be represented as excitation of that particle to a continuum of energy states with an energy spread |$c$>. The strength of interaction between |$d$> and |$c$> is called coupling efficiency.

In Markel's correspondence[9] to Zou and Schatz's work,[7] it was discussed in depth on what the correct representation for the dipole resonance of spheres is. However, in this work, we deal with a different system composed of nanoparticles with rod shape. We choose indium-tin-oxide nanorods (ITONRs) as example, as we have demonstrated in our previous work that we can controllably grow well aligned and ordered ITONR array structure with tunable height, width, and pitch size. Furthermore, we demonstrated very interesting plasmonic resonances in the infrared.[10-13] Due to the elongated geometry of the nanorods, they have different resonances depending on the excitation electric field polarization with respect to the axial axis of the nanorods. When the electric field is along the radial axis of the rods, a transverse plasmonic resonance ($t$-LSPR) occurs, while the electric field along the axial axis excites a longitudinal plasmonic resonance ($l$-LSPR).[10] $t$-LSPR is very sensitive to the shape of the cross section, which is difficult to fit with an analytical model in ITONRs case since the single crystalline ITONR has a square-shaped cross section. On the other hand, $l$-LSPR is not sensitive to the fine-features of the nanorods, and they can be varied across a very large range by changing the aspect ratio, as shown in **Figure 2**. Therefore, $l$-LSPR provides the fundamental advantage over spheres given its vastly tunable resonance and thus the ease of achieving photonic-dipole coupling at arbitrary wavelength.

We provide a simple analytical dipole resonance fitting for the *l*-LSPR resonance and compare the fitting with the finite element simulation result. The polarizability of each ITO nanorod can be expressed as,

$$\alpha = -\frac{\delta}{\omega^2 - (\omega_{LSPR})^2 + i\gamma\omega} \qquad (1)$$

, where $\alpha$ is the polarizability, $\delta$ is the strength of the resonance, $\omega$ is the frequency of the photon ($2\pi c/\lambda$), $\omega_{LSPR}$ is the resonance frequency, and $\gamma$ is the damping factor. Subsequently, the extinction cross-section can be calculated based on the well-known formula,[14]

$$\sigma_e = Im(4\pi k\alpha) \qquad (2)$$

, where $\sigma_e$ is the extinction cross-section and $k$ is the wavevector of the incident photon.

In order to verify the validity of the fitting model, we have simulated the response of a typical isolated ITO nanorod (size: 185 nm x 185 nm x 1 μm), with the light incident from the side and polarized along the axial direction of the nanorod, and compared it with our fitting, as shown in **Figure 3**. A good match between simulation and experiment has been obtained.

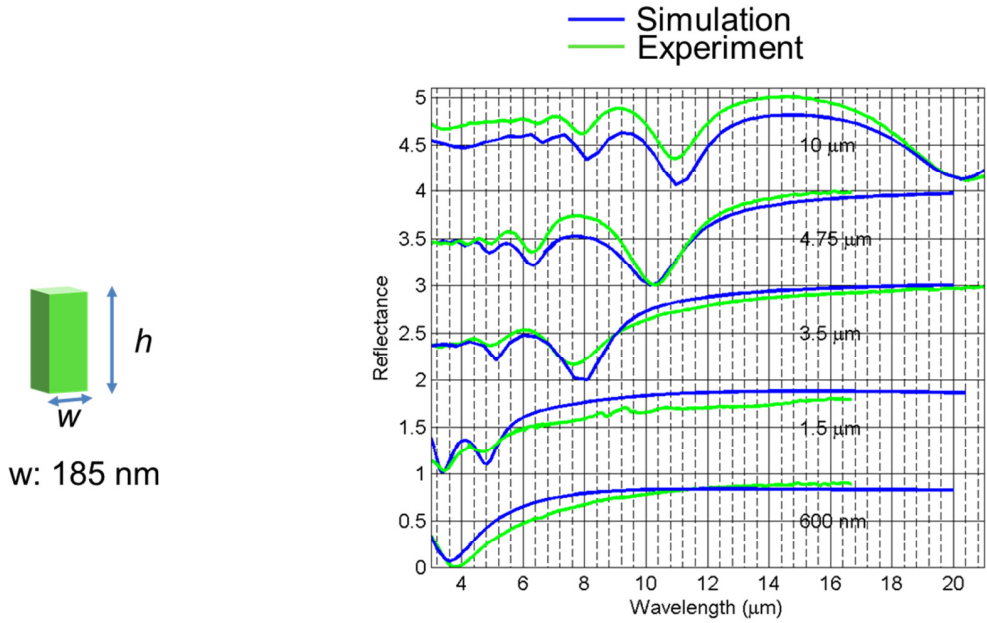

**Figure 2. Demonstration of the tunable l-LSPR for ITO nanorod arrays. The nanorods are vertically aligned on the substrate. They form square lattice with 1.2 μm lattice spacing. A series of nanorod arrays with five different heights were synthesized, with other parameters kept the same. The cross-section of the nanorods is square, due to the natural formation of facets of single crystalline nanorods.[12] An ITO thin film of 200 nm thickness was deposited before the array growth. It is a reflective layer enabling the reflectance measurements. The reflectance dips are the resonance positions. For longer nanorods, more than one oscillation can be seen and they are the higher order resonances. Common parameters of the system: side width (w) of the nanorods is 185 nm. Plasma frequency ($\omega_p$) = 1.55 eV and damping factor ($\gamma$) = 0.062 eV. High frequency dielectric constant ($\varepsilon_{inf}$) = 3.95.**

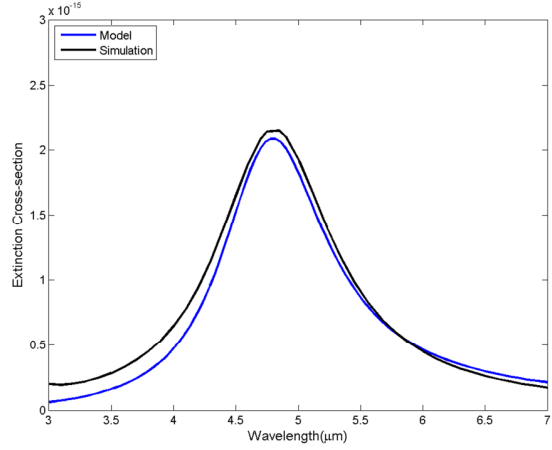

**Figure 3. Comparison with the fitting model used in this work and FEM simulation result based on real structure.**

With the formula for polarizability determined, we proceed to apply the lattice summation method.[7] In the nanorod array structure, the rods are periodic and vertically aligned, as shown in **Figure 4**. Furthermore, the radiated field from the *l*-LSPR of each rod propagates in the direction normal to the axial axis of the rod. Hence, the radiation field from each nanorod strongly interacts with its neighboring, which potentially leads to much stronger electromagnetic coupling among the nanorods than the cases in which the rods all lie in the plane.[15]

In the coupled dipole model, the interaction among the dipoles of the nanorod is taken into account with a structure factor $S(a,b)$, which is a function of the lattice parameter shown in **Figure 4**. **Equation (2)** is modified to,

$$\sigma_e \sim Im(\frac{4\pi k\alpha}{1-\alpha S(a,b)}) \qquad (3)$$

As we discussed above, *l*-LSPR is excited with electric field along the axial axis of the nanorods. Therefore, the maximum excitation occurs when the incident photon has the wavevector *k* in the plane of the 2-dimensional array (see **Figure 4**). We also pick the direction of wavevector to be along one the principal axis *a* in this study. The structure factor $S(a,b)$ can then be written as,

$$S = \sum_{m,n\neq 0} \left[\frac{(ikr_{mn}-1)e^{ikr_{mn}}}{r_{mn}^3} + \frac{k^2 e^{ikr_{mn}}}{r_{mn}}\right] e^{i\vec{k}\cdot\vec{r}_{mn}} \qquad (4)$$

$$\vec{r}_{mn} = m\cdot a\hat{\imath} + n\cdot b\hat{\jmath} \qquad (5)$$

, where *m* and *n* are the integers 0, ±1, ±2, ±3,…, $\hat{\imath}$ and $\hat{\jmath}$ are the elemental vector along the two principal axis of the lattice, $\vec{r}_{mn}$ is the vector pointing from dipole located at (*m*, *n*) to the dipole of interest at (0, 0).

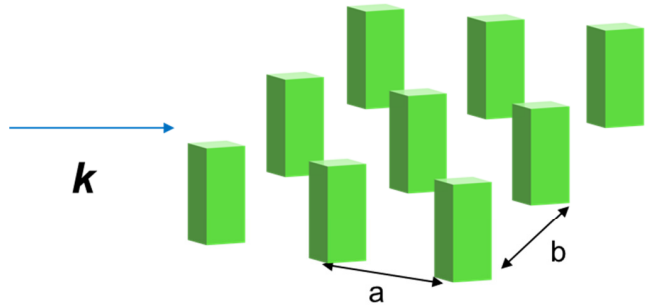

**Figure 4. A schematic diagram showing the interaction of photon with the nanorod array. a and b and lattice spacings along the two principal axes of the array.**

In this analytical model outlined above, only $\delta$, $\omega_{LSPR}$, and $\gamma$ in **Equation (1)** need to be fitted. We next simulated an array of nanorods with geometry shown in **Figure 4**, with square lattice ($a = b = 5$ μm), h = 1 μm, w = 150 nm. From experimental perspective, it is difficult to measure the interaction of array with photons incident 90° to the normal of the array plane. Hence, in the simulation, the incident wave was launched at an incident angle of 80°. The extinction curves from both isolated nanorod and nanorod array are plotted in **Figure 5**. All the features from the simulations are well captured in this simple analytical model. The photonic modes located at 3 μm, 3.5 μm, and 4.2 μm are red-shifted with respect to the simulation. This is because the wavevector of the incident light is assumed to be normal to the plane of the array while in the simulation the angle is less than normal.

We also plotted the near field intensity distribution around the nanorod at the peaks of resonance for both isolated nanorod and nanorod in an array, in the inset of **Figure 5**(b). The near field enhancement in the array is 2-orders of magnitude higher than the isolated nanorod. Furthermore, the width of the main peak is significantly narrower than that from the pure *l*-LSPR peak. The Full-width-half-maximum (FWHM) of the main resonance in the array is 1/40 of that from pure *l*-LSPR (see **Figure 5**(b)).

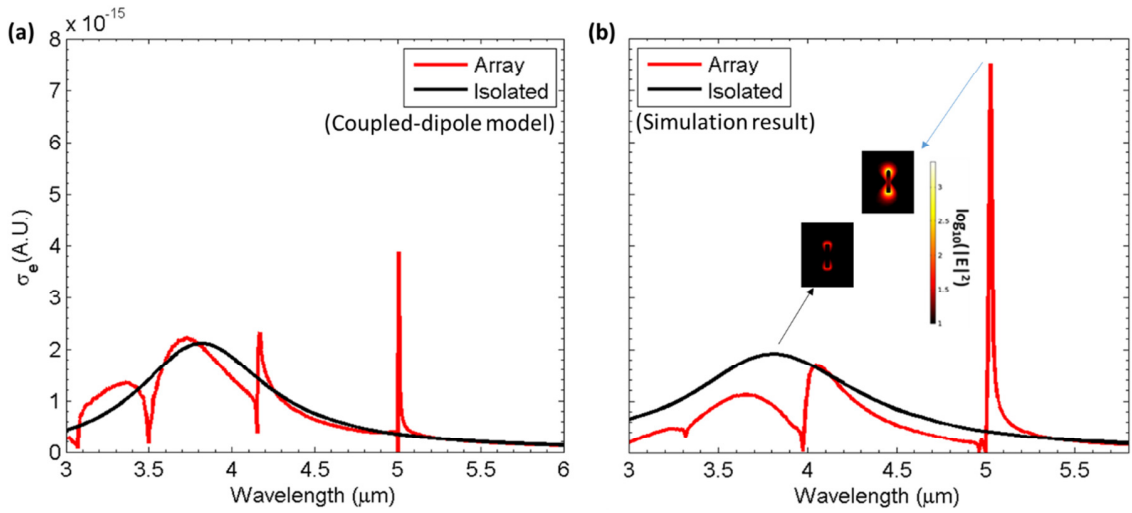

**Figure 5. Comparison between the analytical model and the simulation result.** $\delta = 10^{-16}$, $\omega_{LSPR} = 0.325$ eV, and $\gamma = 0.08$ eV. The insets in (b) show that the near field profiles of the plasmonic resonance of a pure *l*-LSPR mode from the black curve and the coupled resonance from the red curve.

In both the simulation and analytical model, we have not discussed the effect of substrates. However, substrates usually have very strong effect on optical resonances, especially the sharp resonance of interest shown in **Figure 4**. This can be attributed to the strongly dependence of resonance on the refractive index of the surrounding material. When the upper boundary condition of the nanorod (which is air) is different from the lower boundary condition (which is the substrate), the offset between the resonant frequencies upward and downward (denoted by ω₁ and ω₂ in **Figure 6**(a)) strongly quenches the resonance peak of plasmonic-photonic coupling. This can be clearly seen in **Figure 6**(b), in which we plotted the simulation result of the extinction cross section of a freestanding nanorod array and a nanorod array on the substrate. The remedy for recovering the sharp resonances that various research groups have demonstrated is to embed the entire array in an index-matching fluid which has the same refractive index as the substrate, so a homogeneous environment refractive index is preserved.[16, 17]

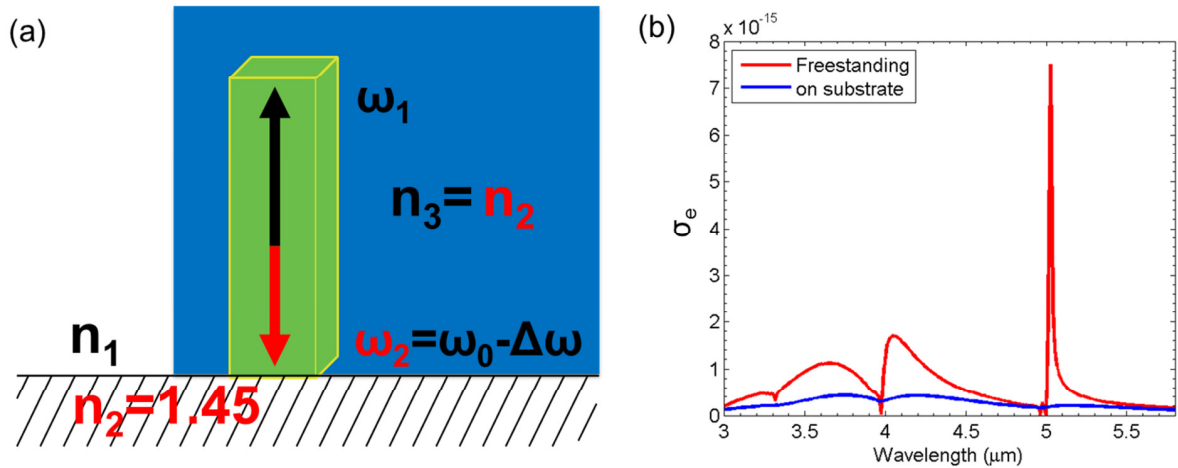

**Figure 6. (a) The schematic diagram showing the damping effect of substrate due to the refractive index differences. (b) The simulation results of both freestanding array and array on the substrate with n = 1.45.**

To recover the sharp resonance without index-matching so that the field-enhanced regime can be easily accessed for different applications, we resort to an alternative method – replacing the dipole resonance with a monopole resonance. It is known in radio-frequency applications that the image theory has been applied to transform a half-wavelength antenna to a quarter-wavelength antenna by using a good conductive surface (which can be approximated as perfect electric conductor (PEC)). Similarly, a vertical dipole sitting on PEC surface is equivalent to a freestanding dipole with twice the dipole strength. This concept should apply equally well in the infrared regime if a good conductive surface can be found.

Gold has a very short skin-depth (around 20 nm) in the infrared regime. By using gold as our PEC (shown in **Figure 7**), we have successfully demonstrate the quarter-wavelength monopole, which has the equivalent spectral features shown in **Figure 6** with freestanding array. The detailed analysis together with experimental realization is presented elsewhere.[4]

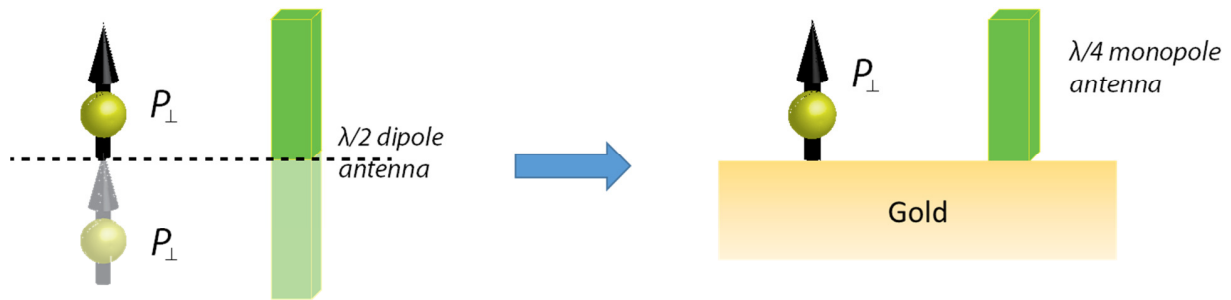

**Figure 7. A half-wavelength antenna can be transformed into a quarter-wavelength antenna by using a perfect conductive substrate. Gold is a very good conductor, which can be treated approximately as perfect electric conductor in the infrared region.**

## 3. CONCLUSIONS AND FUTURE WORK

In this report, we have analyzed the validity of coupled-dipole model in describing vertically aligned and ordered ITO nanorod array and obtained very good match between the analytical model and finite element simulations. Specifically, sharp resonances due to photonic-plasmonic coupling were observed both in the analytical model and simulation. We presented some preliminary results showing the tunability of *l*-LSPR, as it is a mode strongly dependent on the aspect ratio of the nanorods, which can be controlled experimentally. This tunable *l*-LSPR is critical for engineering the high-Q resonances at an arbitrary wavelength. The experimental and simulation results further show the good approximation of dipole resonance with the proposed fitting model, due to the fact that *l*-LSPR is insensitive to the fine features of nanorods. At the end of the discussion, we have outlined the procedure of how to mitigate the index-matching problems by image theory, with detailed discussion presented elsewhere.[4] The high-Q, high field-enhancement resonances discussed here can be engineered easily by change the lattice parameters and *l*-LSPR position, which is a versatile system for a wide-range of applications requiring a flexibly-engineered resonance.

**Acknowledgments:**

Supported by NSF funding (DMR-1121262 and DMR 0843962), QUEST computational resources (Project p20194 and Project p20447) and Center for Nanoscale Materials in Argonne National Laboratory (Project CNM 25883 and Project CNM 30831). Various characterizations were done in NUANCE center and KECK II Facilities in Northwestern University. The NUANCE Center and KECK II Facilities are supported by the NSF-NSEC, NSF-MRSEC, Keck Foundation, the State of Illinois, and Northwestern University. E-beam lithography was performed with JEOL-9300 in the Center for Nanoscale materials at Argonne National Laboratory. Use of the Center for Nanoscale Materials was supported by the U. S. Department of Energy, Office of Science, Office of Basic Energy Sciences, under Contract No. DE-AC02-06CH11357.